# Transport Properties of Graphene Nanoribbon Transistors on Chemical-Vapor-Deposition Grown Wafer-Scale Graphene

Wan Sik Hwang[1, a)], Kristof Tahy[1], Xuesong Li[2], Huili (Grace) Xing[1], Alan C. Seabaugh[1], Chun Yung Sung[2], and Debdeep Jena[1, b]

[1]Department of Electrical Engineering, University of Notre Dame, Notre Dame, IN 46556, USA

[2] IBM T. J. Watson Research Center, Yorktown Heights, NY10598, USA

a) Electronic mail: whwang1@nd.edu and b) Electronic mail: djena@nd.edu

**ABSTRACT**

Graphene nanoribbon (GNR) field-effect transistors (FETs) with widths down to 12 nm have been fabricated by electron beam lithography using a wafer-scale chemical vapor deposition (CVD) process to form the graphene. The GNR FETs show drain-current modulation of approximately 10 at 300 K, increasing to nearly $10^6$ at 4 K. The strong temperature dependence of the minimum current indicates the opening of a bandgap for CVD-grown GNR-FETs. The extracted bandgap is estimated to be around 0.1 eV by differential conductance methods. This work highlights the development of CVD-grown large-area graphene and demonstrates the opening of a bandgap in nanoribbon transistors.

Graphene, a single sheet of carbon atoms bonded in a hexagonal lattice, is being investigated for its promising electronic properties such as the high mobility of carriers and its ultrathin nature [1, 2]. Though significant progress has been made experimentally to realize graphene devices, much of the work has used single-layer graphene obtained by mechanical exfoliation. For device and circuit level applications, wafer-scale graphene is necessary. Two promising approaches have been developed for large-area graphene growth. The first is the growth of graphene on silicon carbide (SiC), commonly referred to as an epitaxial graphene [3, 4]. The second is the growth of graphene on suitable metals by chemical vapor deposition (CVD), followed by a transfer process onto any substrate [5]. Both methods are attractive, because graphene size is only limited by the substrate materials such as SiC for epitaxial method or metal for CVD method respectively. Though the latter approach involves a transfer process, it is relatively inexpensive and offers the freedom to transfer the resulting graphene onto suitable substrates. Graphene devices have been extensively reported on both exfoliated and epitaxial graphene [3, 5]. Recently, CVD-grown two dimensional (2D) graphene devices for RF applications have been reported [6]. however, there are no reports of the properties of graphene nanoribbons formed on large-area grown CVD graphene, which is the subject of this paper.

The opening of energy gaps in 2D graphene by quantum confinement is highly desirable for electronic switching, both for conventional FETs as well as for tunnel FETs (TFETs) [7, 8]. Experimental observations of such gaps have been reported for GNRs patterned from exfoliated graphene [9, 10], but not on wafer-scale graphene. In this work, we report the fabrication and electronic properties of GNR transistors on CVD-grown large-area graphene. In particular, we observe the opening of a bandgap as large as 0.1eV in GNR-FETs as measured by a differential conductance method [9].

CVD graphene was grown and transferred onto $t_{OX}$ = 90 nm SiO$_2$/p$^+$ Si substrates [11]. Hydrogen silsesquioxane (HSQ) diluted with methylisobutylketone (MIBK) was used as an e-beam resist to form GNRs with width $W$ = 12 nm. Details of the HSQ process have been discussed [12]. Source/drain contact metals Cr/Au (5/100 nm) were deposited by electron-beam evaporation. Figure 1 (a) shows the schematic device structure and layout in top and cross-sectional views. The GNR is connected to the 2D graphene by a linear lithographic flare from the width of 12 nm to the tens-of-microns scale. The source and drain contact metals sit on top of the 2D graphene regions. The graphene is covered by HSQ as also shown in the cross-section view of Fig. 1(a). Figure 1 (b) shows a scanning electron microscope (SEM) image of the GNR region in the device. The inset image enlarges the GNR region and reveals that the width of GNR is 12 nm and this is a measure of the HSQ and it is not known exactly how this pattern roughness transfers to the GNRs. From the SEM image, the nanoribbon edge appears relatively smooth.

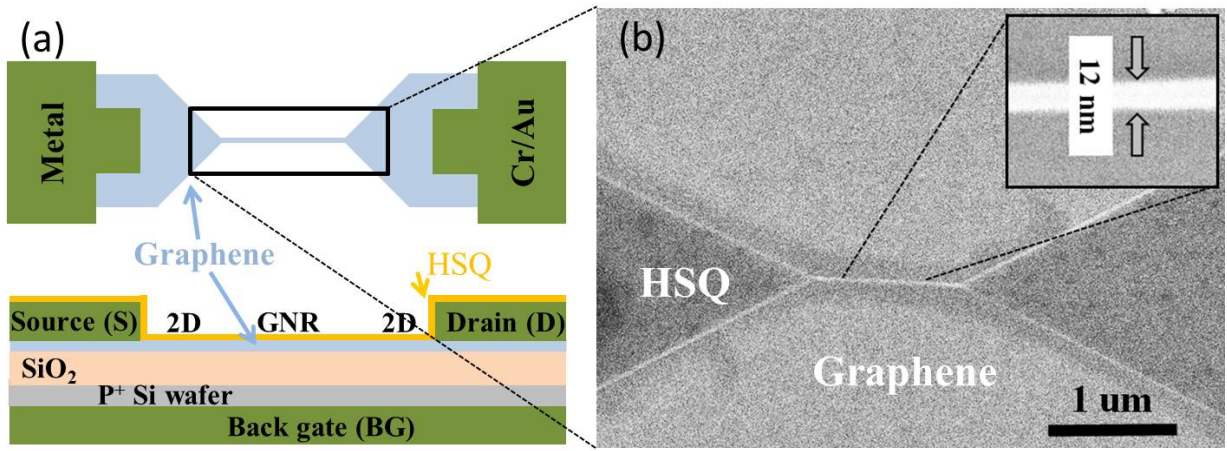

**FIG**. 1. (a) Schematic device structure and layout of the back-gated GNR FET. (b) SEM image of the GNR with an inset showing a magnified view of the nanoribbon.

Electrical measurements were performed in vacuum from room temperature (300 K) to 4 K. Figure 2 (a) shows the measured drain current $I_D$ versus the back-gate voltage $V_{BG}$ for a 12 nm wide GNR-FETs at various temperatures. The gate modulation is approximately 10× at 300 K. The gate modulation increases to nearly $10^6$× at 4 K. The strong temperature dependence of the minimum current indicates the opening of a bandgap. The behavior is quite distinct from 2D graphene FETs where the modulation remains essentially unchanged over similar temperature range due the absence of a bandgap [12, 13]. One may estimate the bandgap created by quantum confinement in an armchair-edge GNR to be $E_G \sim 2\pi\hbar v_F / 3W$, where $\hbar$ is the reduced Planck's constant and $v_F \sim 10^8$ cm/s is the Fermi velocity characterizing the conical bandstructure of graphene [14]. Although the measured GNRs likely contain a mixture of armchair and zig-zag edges, we assume the inverse dependence of bandgap on GNR width will not be altered by the nature of the edge. For $W = 12$ nm, we expect $E_G \approx 0.1$ eV. The GNR is connected to the 2D graphene regions through 'adiabatic' contacts, meaning the bandgap should change continuously from 0 eV in the 2D regions to the bandgap of the GNR at the ribbon. Thus the contacts to the GNR are effectively of the Schottky-barrier kind, with height $\phi \sim E_G/2 = 50$ meV. The Fermi-tail of carriers at room temperature is substantial, and thus a large off-state leakage current due to thermionic emission is expected. However, as the temperature is lowered, $kT << \phi$ and thus the off-state current reduces drastically since carriers must now either tunnel through the gap, or hop through 'defect' states. At this stage, we have not attempted a more detailed numerical calculation of the temperature-dependence of the current beyond this qualitative picture, which captures the essence of the experimental observation. Figure 2 (a) also shows ambipolar behavior. Depending on the $V_{BG}$, electrons or holes become main conduction carrier type. The window becomes clearer at low temperatures when the thermionic emission current is suppressed. Figure 2(b) shows a family of $I_D$ - $V_{DS}$ curves for the GNR-FETs at various $V_{BG}$. Figure 2 (a) and (b) clearly shows the substantial modulation at low $V_{DS}$ due to the movement of the Fermi level controlled by the back gate voltage. At $V_{DS}$ substantially larger than the gap, the current modulation is lower, similar to the breakdown behavior in semiconductor FETs. Current densities exceeding 1000 µA/µm are measured at high drain biases, consistent with recent reports for unzipped GNRs [15].

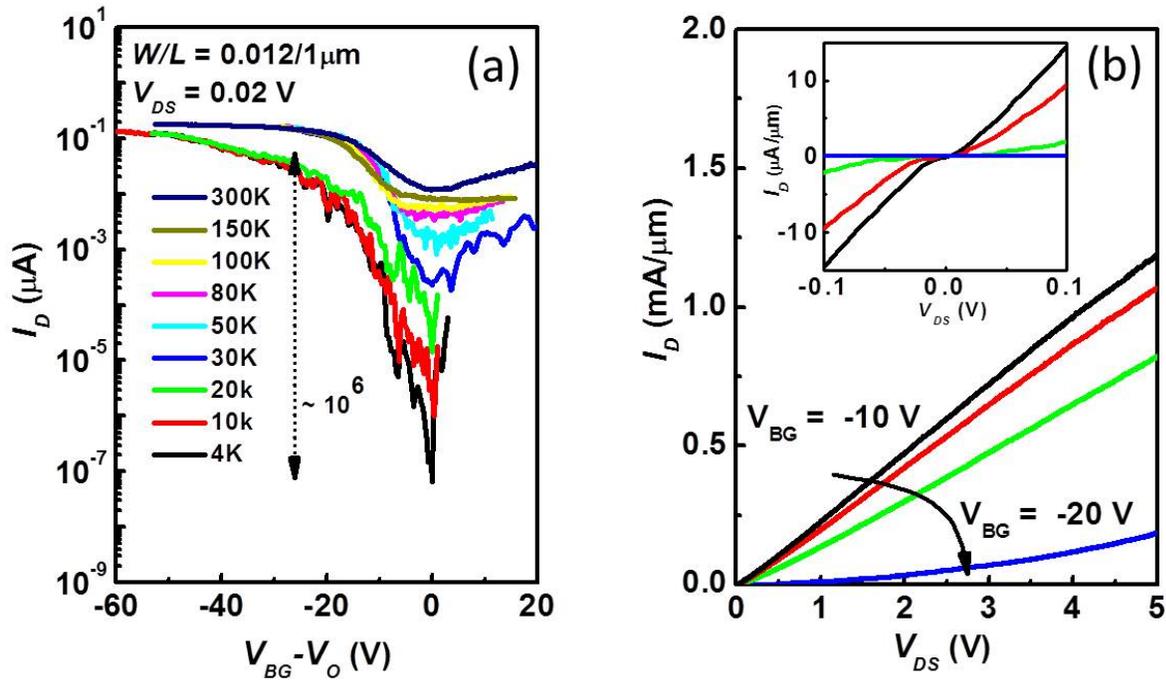

**FIG**. 2. Transport properties of back-gated CVD GNR FET of width 12 nm. (a) Drain current vs. back-gate voltage and temperature. (b) Common-source transistor characteristics at 4 K.

Figure 3 (a) shows the differential conductance log ($dI_{DS}$ / $dV_{DS}$) vs. $V_{DS}$ and $V_{BG}$ for the GNR-FET at 4 K. The differential conductance is represented as a color in logarithmic scale; black (dark) color indicates low conductance and red (light) color indicates high conductance.

The vertical extent of the dark diamond shape is indicative of the GNR gap [9]. The extracted GNR bandgap from this method is around 0.1 eV as discussed earlier, consistent with what is expected [9, 14]. Though the experimental extraction is close to the model, it is worth discussing recent reports on the extraction of bandgap due to transport arising from hopping between quantum-dot like localized regions in GNRs [16, 17]. For GNR widths greater than 40 nm which the band-gap is less than 0.04 eV, extraction of the low bandgap by the conductance method is expected to be strongly affected by background potential disorder, or hopping between localized states. However when the bandgap is substantially larger than the potential disorder, the disorder acts as a weak perturbation, similar to dopants in semiconductors. Since the GNRs fabricated here have smaller width than the previous reports [9, 10, 16, 17], the gap reported is indicative of the modification of the density of states by quantum confinement. Since the differential conductance at low temperature is proportional to the density of states, an energy gap of ~0.1 eV can be also inferred from Fig 3(b).

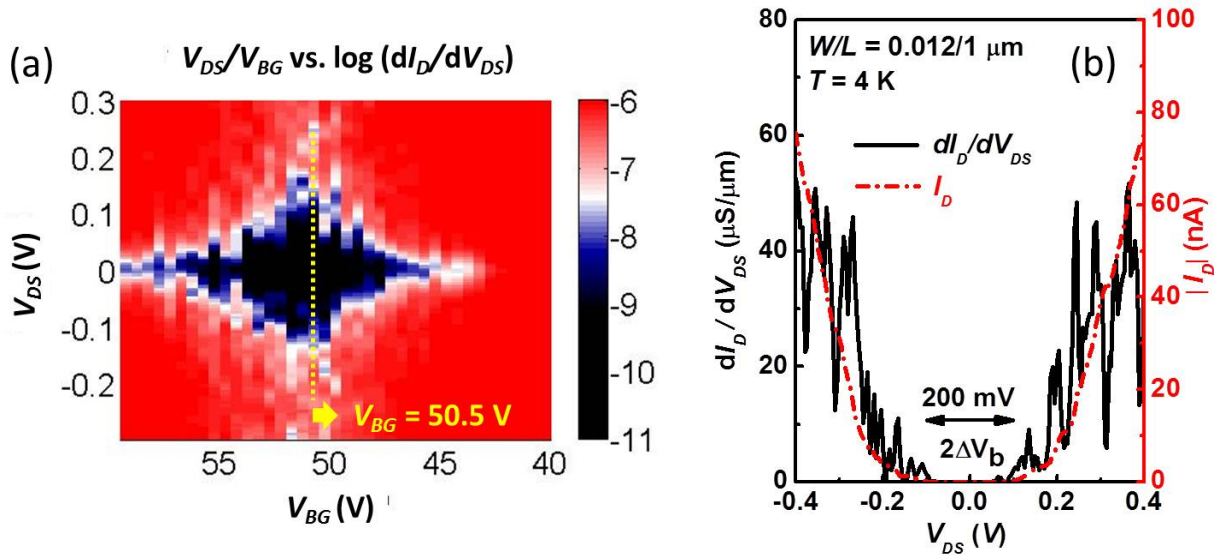

**FIG**. 3. (a) Differential conductance map of a 12 nm GNR FET as a function of $V_{DS}$ and $V_{BG}$ at 4K. (b) Differential conductance and absolute drain current vs. drain-to-source voltage at a back-gate bias of 50.5 V.

A more quantitative theory of the behavior of the GNR-FETs requires knowledge of the precise edge roughness and edge-geometry, which is not available at this point. In spite of this unknown, in this work we show that GNRs fabricated by lithography from large-area CVD graphene demonstrate a substantial bandgap opening which is similar to that from exfoliated graphene. The large-area graphene will thus allow a systematic study of large numbers of similar GNRs, which is currently underway. Statistical analysis of the properties of GNRs of similar widths is expected to move the field of electronics based on graphene nanostructures forward. The experimental result shown here should be considered a first step towards such goals.

**ACKNOWLEDGEMENTS**

This work was supported by the Semiconductor Research Corporation (SRC), Nanoelectronics Research Initiative (NRI) and the National Institute of Standards and Technology (NIST) through the Midwest Institute for Nanoelectronics Discovery (MIND), and by the Office of Naval Research (ONR) and the National Science Foundation (NSF).